\documentclass[runningheads]{llncs}

\usepackage{amsfonts,amsmath, amssymb}

\usepackage{empheq}
\usepackage{mathpartir}
\usepackage{semantic}
\usepackage{diagbox}
\usepackage{listings}
\usepackage{bbding}

\newcommand{\drule}[3] {\inferrule*[left={#1}]{#2}{#3}}

\newcommand{\tsym}[1]{\:\mbox{\texttt{#1}}\:}
\newcommand{\ntsym}[1]{\:\langle\mbox{\emph{#1}}\rangle\:}

\newcommand*\widefbox[1]{\fbox{\hspace{1em}#1\hspace{1em}}}

\newenvironment{bnf}{%
    \scriptsize
    \setkeys{EmphEqEnv}{align*}%
    \setkeys{EmphEqOpt}{box=\widefbox}%
    \EmphEqMainEnv%
}{
    \endEmphEqMainEnv
}

\usepackage[T1]{fontenc}
\usepackage{graphicx}
\usepackage[hyphens]{url}
\begin{document}

\title{Operational Semantics for Crystality: A Smart Contract Language for Parallel EVMs}

\titlerunning{Operational Semantics for Crystality}

\author{Ziyun Xu\inst{1}\orcidID{0009-0003-7824-6502} \and
Hao Wang\inst{2}\orcidID{0000-0003-3557-6301} \and
Meng Sun\inst{1}\orcidID{0000-0001-6550-7396}\inst{ (}\Envelope\inst{)}}

\authorrunning{Z. Xu et al.}

\institute{School of Mathematical Sciences, Peking University, China \and
International Digital Economy Academy (IDEA), China\\
\email{xuziyun@pku.edu.cn, wanghao2020@idea.edu.cn, sunm@pku.edu.cn}}

\maketitle            

\begin{abstract}
The increasing demand for scalable blockchain has driven research into parallel execution models for smart contracts. Crystality is a novel smart contract programming language designed for parallel Ethereum Virtual Machines (EVMs), enabling fine-grained concurrency through Programmable Contract Scopes and Asynchronous Functional Relay. This paper presents the first formal structural operational semantics for Crystality, providing a rigorous framework to reason about its execution. We mechanize the syntax and semantics of Crystality in the theorem-proving assistant Coq, enabling formal verification of correctness properties. As a case study, we verify a simplified token transfer function, demonstrating the applicability of our semantics in ensuring smart contract correctness. Our work lays the foundation for formally verified parallel smart contracts, contributing to the security and scalability of blockchain systems.


\end{abstract}

\section{Introduction}
Blockchain technology has revolutionized decentralized computing by providing an immutable, transparent, and trustless execution environment  \cite{Bitcoin}. At the heart of blockchain ecosystems are smart contracts-self-executing programs that automate decentralized agreements without relying on intermediaries. However, current mainstream blockchain platforms like Ethereum are built upon inherently sequential execution models, which severely limit transaction throughput and scalability \cite{Georgiadis}. As blockchain adoption grows, with increasing numbers of accounts, transactions, and decentralized applications (DApps), these execution inefficiencies create performance bottlenecks that hinder real-world usability\cite{sanka2021systematic}.

This challenge has motivated extensive research into parallel execution mechanisms aimed at enhancing blockchain efficiency and scalability while preserving decentralization and security. Parallel Virtual Machine (Parallel VM) techniques \cite{Solana,Aptos,Sei,Sui,Monad} have emerged as a promising solution \cite{anjana2019efficient,chen2021forerunner,garamvolgyi2022utilizing,gelashvili2023block,herlihy2008transactional,qi2023smart,saraph2019empirical}, allowing concurrent execution of smart contracts across multiple virtual machines (VMs). While these approaches have achieved orders of magnitude improvements in transactions per second (TPS), they are 
constrained by three fundamental issues:
\begin{itemize}
\item Inefficient Concurrency Control: Optimistic concurrency control \cite{Aptos,Monad,kung1981optimistic} suffers from excessive rollbacks due to contention on shared states, while pessimistic concurrency control \cite{Solana,Sui} incurs high overhead due to dependency analysis and locking.
\item Failure to Exploit Commutativity: Most parallel execution frameworks strictly enforce block-defined transaction ordering, which disregards the commutativity of many smart contract operations, such as token transfers, voting \cite{voting}, and airdrops \cite{airdrop}, that could otherwise be executed in parallel. This limitation stems from their coarse-grained, transaction-level execution paradigm.
\item Lack of Language-Level Parallelism Constructs: Smart contract languages like Solidity \cite{Solidity} and Move \cite{Move} operate under a shared-everything model without built-in mechanisms to express fine-grained concurrency at the state level. This limits their ability to fully exploit parallel execution opportunities.
\end{itemize}

In response to these challenges, Crystality, a novel smart contract programming model recently proposed by Wang et al. \cite{Crystality2025}, introduces an 
approach to facilitate parallel execution across multiple EVMs. It provides a fine-grained state-level parallelism mechanism at the programming language level. Crystality introduces Programmable Contract Scopes as directives to partition contract states into independent, non-overlapping segments, and to decompose contract functions into smaller, parallelizable components based on state dependencies. This enables programmers to explicitly express state-level parallelism while allowing the underlying system to enforce asynchronous execution. Another key innovation in Crystality is Asynchronous Functional Relay, which orchestrates execution flow across multiple VMs based on data dependencies, thereby enabling non-blocking execution of commutative operations. Experimental evaluations have demonstrated that Crystality significantly improves blockchain throughput and scalability without compromising decentralization or security \cite{Crystality2025}.

Despite its promising results, Crystality currently lacks a formal semantics \cite{WG}, a critical prerequisite for rigorously defining program behavior, analyzing concurrency properties, and providing correctness guarantees \cite{HH}, especially in high-assurance domains such as financial transactions, supply chain management, and decentralized governance. This is particularly important given its novel execution paradigm, where state partitioning, asynchronous relays, and cross-VM execution introduce intricate interactions that require precise formal treatment. A structural operational semantics \cite{Plotkin} for Crystality would bridge the gap between its high-level programming constructs and their low-level execution semantics on parallel EVMs, ensuring Correctness-by-Construction.

While formal semantics for smart contract languages has received considerable attention, existing work has primarily focused on languages based on sequential execution models, particularly Solidity \cite{crosaratowards,Hajdu1,Hajdu2,hildenbrandt2018kevm,jiao2018,jiao2020semantic,Marmsoler,Yang2020,zakrzewski2018towards}. These semantics are not directly applicable to Crystality, as it introduces new execution paradigms and features that require novel formalization techniques.

In this paper, we present the first formal semantics for Crystality, 
making the following key contributions:
\begin{itemize}
\item {\bf Structural Operational Semantics for Crystality:}
We present a mathematically precise execution model for Crystality, including formal definitions of Programmable Contract Scopes and Asynchronous Functional Relay.
To the best of our knowledge, this is the first formal semantics for a smart contract language explicitly designed for parallel EVMs.

\item {\bf Formalization and Verification in Coq:}
We implement Crystality’s core syntax and semantics in Coq, establishing a foundation for mechanized formal reasoning and verification.
As a case study, we verify a simplified Crystality token transfer function, demonstrating the practical capabilities of our framework for verifying smart contract correctness.
\end{itemize}

The remainder of this paper is organized as follows:
Section 2 provides an overview of Crystality, including its core constructs, syntax, and a motivating example.
Section 3 introduces the structural operational semantics, detailing its formalization and execution rules.
Section 4 presents our Coq-based formalization and verification case study.
Section 5 briefly discusses related work.
Section 6 concludes the paper and discusses future research directions.

\section{A Primer on Crystality }
Crystality is a novel parallel programming model designed for smart contracts on parallel 
EVMs. It allows each EVM to independently manage a portion of the ledger state and transactions, operate in parallel, and coordinate via a relay mechanism. The ledger state maps addresses to account states (key-value pairs), while transactions, as cryptographically signed messages, modify states via smart contract function calls.

Smart contracts are self-executing programs which maintain the ledger state that can be modified through function calls triggered by transactions. A smart contract comprises:
\begin{itemize}
\item State Variables: Key-value pairs storing contract data, such as account balances.
\item Functions: Operations modifying state variables, such as transferring tokens.
\end{itemize}

Crystality facilitates fine-grained parallelism through two core mechanisms: Programmable Contract Scopes and Asynchronous Functional Relay.

\subsection{Programmable Contract Scopes}
Each contract variable and function is annotated with a \texttt{@scope} directive, defining its execution boundary. Scopes partition contract states into isolated execution units, minimizing conflicts and ensuring parallelism. This design enables concurrent state updates in Crystality.

For state variables, scopes are the storage regions in which the variables reside. The storage for state variables (the ledger state) is explicitly and deterministically partitioned into distinct, non-overlapping regions across multiple execution engines, ensuring balanced workload and efficient resource utilization. The scope of a state variable also defines how many instances of the variable exist on the blockchain.

There are three main types of scopes:

\begin{bnf}
    \ntsym{scope} ::= \tsym{@address} | \tsym{@engine} | \tsym{@global}
\end{bnf}

\begin{itemize}
\item Address Scope (\texttt{@address}): 
This scope partitions contract states based on user addresses, where each execution engine manages a subset of addresses. Each variable within this scope has one instance per valid address on each engine. 
\item Engine Scope (\texttt{@engine}): 
This scope is specific to each execution engine, representing local-shared states that can be immediately read or written by functions executed within the same engine. Each variable within this scope has a single instance per engine.
\item Global Scope (\texttt{@global}):
This is a logically singleton scope accessible by all execution engines. Variables in this scope are globally readable, but their updates require synchronization across all engines. It requires all execution engines to perform the same update operation in order to modify the value of the global state variable. Each variable within this scope has only one instance across the entire blockchain.
\end{itemize}

In Crystality, a function's scope determines its access to state variables. Unlike Solidity, function invocation requires specifying the current scope. A function can directly access variables and functions within its own scope and read state variables if their scope is equal to or broader than the function’s. Write access follows the same rule, except \texttt{@global} state variables, which only \texttt{@global} functions can modify. Table~\ref{tab1} summarizes these rules, where `R' denotes read permission, `W' denotes write permission, and `-' indicates no access.

\begin{table}
\centering
\caption{Functions Access Permissions to State Variables}\label{tab1}
\begin{tabular}{|l|l|l|l|}
\hline
\diagbox{Func. Scope}{State Var. Scope} &\texttt{@address}&\texttt{@engine}&\texttt{@global}\\
\hline
\texttt{@address} & R W & R W & R\\
\texttt{@engine}  & - & R W & R\\
\texttt{@global}  & - & - & R W\\  
\hline
\end{tabular}
\end{table}
In addition to state variables, Crystality supports temporary variables within functions. These are local to functions within specific scopes and stored in partitions corresponding to their scope. However, we can conceptualize temporary variables in the same manner as those in traditional programming languages, viewing them as existing in the function call stack.

\subsection{Asynchronous Functional Relay}
A fundamental challenge in parallel computing is managing code execution with respect to data dependencies. Since the required data for contract functions may reside in different scopes, Crystality adopts a unique execution model where functions can invoke one another asynchronously across scopes. In other words, functions can trigger relay calls. This model avoids the need to collect all relevant data into a single scope, enabling transaction routing directly to the scope that holds the relevant data. This approach allows inter-scope communication and facilitates complex workflows that span multiple scopes.

The relay mechanism decomposes transaction execution into multiple invocations of fine-grained, scope-narrowed functions. Functions in different scopes can execute in parallel without blocking each other. State updates are pipelined and concurrent across different execution engines.

A relay call is essentially a message containing the target scope, the identifier of the function to be invoked, and the associated arguments to pass. 

\begin{bnf}
    \ntsym{relaycall} ::= ( \tsym{relay @}\ntsym{exp} | \tsym{relay @engines} | \tsym{relay @global} ) \ntsym{identifier} \tsym{(} \ntsym{exp}^{*} \tsym{)} \tsym{;}
\end{bnf}

There are three types of relay target. The first type is used when relaying a function to a specific valid address. The value of \emph{exp} is the target address. The second and third type broadcast to all engines. The difference between the second and third types lies in the scope of the function being relayed. The scope of the function must match the scope of the relay target. For instance, if the relay is directed to the \texttt{@global} scope, the function's scope must also be \texttt{@global}.

Crystality deliberately avoids relaying to a specific engine to maintain scalability and abstraction from blockchain implementation details.

The relay call is converted to a relay transaction by the execution engine and sent to the destination engine. While a transaction is being executed on the current engine, all dynamically generated relay calls are collected along with their invocation data, which are then packaged into relay transactions, and propagated to the target engine's broadcast network, where they reside in the mempool (the transaction pool that temporarily stores the transactions), awaiting inclusion, confirmation, and execution.

Unlike traditional function calls, where execution and state updates are immediate, a relay call returns instantly, with the actual execution taking place later, 
often in a different scope.

\subsection{Syntax of Crystality}
In the following, we introduce the syntax of Crystality, represented by a variant of Extended Backus-Naur Form (known as EBNF) where:

\begin{itemize}
 \item Terminal symbols are written in \tsym{monospaced fonts}.
 \item Non-terminal productions are encapsulated in  $\ntsym{angle brackets}$.
 \item Zero or one occurence is denoted by $^{?}$, zero or more occurences is denoted by $^{*}$.
\end{itemize}

A contract takes the following syntactic form:
 
\begin{bnf}
    \ntsym{contractDecl} ::= \tsym{contract} \ntsym{identifier} \tsym{\{} \ntsym{statevarDecl} ^{*} \ntsym{funcDecl}^{*} \tsym{\}}
\end{bnf}

\emph{Identifier} serves as a unique name for the contract. \emph{StatevarDecl} defines the various state variables associated with the contract, and \emph{funcDecl} specifies the functions within the contract.

\begin{bnf}
    \ntsym{statevarDecl} ::= \ntsym{type} \ntsym{scope} \ntsym{identifier} \tsym{;}
\end{bnf}

\emph{StatevarDecl} specifies both the type and the scope of the state variable.

The abstract syntax tree of functions is as follows.

\begin{bnf}
    \ntsym{funcDecl} ::=& \tsym{function}  \ntsym{identifier} \ntsym{funcPara} \ntsym{scope} \tsym{returns} \ntsym{type}^{?} \ntsym{funcBody} \\
    \ntsym{funcPara} ::=&  \tsym{(} (\ntsym{type}   \ntsym{identifier} \tsym{,} )^{*} \ntsym{type}   \ntsym{identifier} \tsym{)}  | \tsym{(} \tsym{)} \\
    \ntsym{funcBody} ::=& \tsym{\{} \ntsym{stmt} \tsym{\}} \\
    \ntsym{stmt} ::=& \ntsym{pstmt} | \ntsym{stmt} \ntsym{stmt} | \tsym{if} \tsym{(} \ntsym{exp} \tsym{)} \tsym{then} \tsym{\{} \ntsym{stmt} \tsym{\}} \tsym{else} \tsym{\{} \ntsym{stmt} \tsym{\}} \\ & | \tsym{while}  \tsym{(} \ntsym{exp} \tsym{)} \tsym{\{} \ntsym{stmt} \tsym{\}} \\
    \ntsym{pstmt} ::=& \ntsym{type} \ntsym{identifier} \tsym{;} | \tsym{skip} | \ntsym{identifier} := \ntsym{exp} \tsym{;} | \ntsym{relaycall} \\ & | \ntsym{returnstmt} | \ntsym{identifier} \tsym{(} \ntsym{exp} ^{*} \tsym{)} \\
    \ntsym{exp} ::=& \ntsym{identifier} | \ntsym{identifier} \tsym{(} \ntsym{exp} ^{*} \tsym{)} 
\end{bnf}

Basically, a function declaration consists of the following parts: its identifier, parameters along with their corresponding types, its scope, return type (if applicable), and the function body. The function body is enclosed within braces and consists of some statements. The statement may be primitive statement, sequential composition, conditional, or loop. A primitive statement may be temporary variable declaration, skip, assignment, relay call, return statement, or function call. The expression can be identifier or function call.

The following is an ERC20 token transfer contract in Crystality. The balance variable tracks user balances, and the transfer function moves tokens from the sender to a payee if the sender has sufficient funds.
\begin{lstlisting}[label={lst:Crystality_example}]
contract MyToken {
    uint256 @address balance;
    function transfer(address payee, uint256 amount)
    @address returns
    {
        if (amount <= balance) then {
            balance := balance - amount;
            relay @ payee mint (amount);     
        } else { skip }
    }
}
\end{lstlisting}

User balances are partitioned by address within the \texttt{@address} scope, ensuring each user has a unique balance instance. The transfer function, also scoped to \texttt{@address}, is invoked via a transaction specifying the sender’s address. Execution occurs in the sender's partition, and if the payee's balance is in a different partition, a relay call handles the cross-partition update. This mechanism decouples payer and payee operations, reducing transaction contention.

To proceed with the deposit to the payee, a relay call is initiated with the payee’s address as the target scope, directing execution to the engine managing the payee’s state. The predefined mint function adds the specified amount to the balance.

\section{Structural Operational Semantics of Crystality} 
In this section, we present the structural operational semantics of Crystality. We reveal the idea of the semantics from a general as well as abstract perspective, focusing on the theoretical underpinnings of the language rather than its specific implementation. Our semantics is independent of any particular realization of Crystality, aiming to capture the core behavior and rules governing the language in a formal and generalized manner.

First, we introduce the notations being used in the semantics and explain the necessity of these notations. Following this, we provide the semantics for statement execution, expression evaluation, and transaction execution respectively. 

\subsection{Notations}
In the following, we use $n$ to represent the number of engines and $k$ to represent the number of addresses per engine. We use $\mathbb{A}$ to denote all memory addresses and $\mathbb{B}$ to denote bytes. Then a function $\mathbb{A} -> \mathbb{B}$ can represent storage, where it maps memory addresses to the corresponding byte values. We use $\mathbb{ID}$ to denote the set of all variable names (identifiers) and $\mathbb{IDF}$ the set of all function names (identifiers). Additionally, $\mathbb{T}$ is the set of all variable types.

We use $(\sigma_1,\Omega_1,Prog_1,\sigma_2,\Omega_2,Prog_2,...,\sigma_n,\Omega_n,Prog_n,G)$ to represent the configuration of the entire blockchain system. Within this configuration, $G : \mathbb{A} -> \mathbb{B}$ represents the \texttt{@global} state variable storage. Each $\sigma_i = (\Psi_i,M_i)$ represents the state of the $i$-th engine. $\Omega_i$ represents the mempool of the $i$-th engine, which is an abstraction of the actual mempool maintained by every node that participates in the $i$-th engine. Nodes only store transactions for their own engine, so each $\Omega_i$ is different from one another. Relay calls, once packaged into relay transactions, are sent to $\Omega_i$ for storage. $Prog_i$ represents the program statements that is about to be executed on the $i$-th engine.

Specifically, $\Psi_i = (\Psi_{i,1},\Psi_{i,2},...,\Psi_{i,k},\Psi_{i,s})$ represents the state variable storage for the $i$-th engine, where $\Psi_{i,j}: \mathbb{A} -> \mathbb{B}$ denotes the storage for \texttt{@address} state variables at the  $j$-th address in the $i$-th engine for all $1 \le j \le k$, and $\Psi_{i,s}: \mathbb{A} -> \mathbb{B}$ represents the storage for \texttt{@engine} state variables in the $i$-th engine.

$M_i$ refers to the memory for temporary variables in the $i$-th engine, 
which contains a stack to model new scopes when calling a function. Each layer of the stack is a function $\mathbb{A} -> \mathbb{B}$, representing the memory for temporary variables within the current function scope. $top(M_i)$ returns the top layer of the stack, indicating the memory available in the current function scope. $X = pop(M_i)$ removes the top layer of the stack and returns the new stack. $Y = push(M_i,Z)$ pushes a new top $Z$ to $M_i$, and the result is $Y$. Each layer of $M_i$ is associated with two values: $scope$, which indicates the scope of the function, and $rt$, which represents the return value of the function (if the function has one). Both of these associated values must be defined when the layer is pushed onto the top of the stack. This ensures that, in addition to storing temporary variables, each function invocation layer maintains important contextual information including the current scope and, if applicable, its return value.

In addition, every storage or memory $f: {\mathbb{A}} -> \mathbb{B}$ is associated with a name space $N_f: {\mathbb{ID}} -> {\mathbb{A}} \cup \mathit{None}$ and a type space $T_f : {\mathbb{ID}} -> \mathbb{T} \cup \mathit{None}$, respectively mapping variable identifiers to memory addresses and types. When the identifier $id$ is mapped to $\mathit{None}$ in the name space or type space, it means that $id$ is not defined yet. To access locations in storage or memory, we use $[addr]_{store}^{size}$ to represent the value
stored in position $addr$ of $size$ bytes in $store$. For example, given a variable identifier $id$ allocated in $f$, the notation $[N_f(id)]_{f}^{size(T_f(id))}$ represents the value of $id$. The  size of type $t$ is given by $size(t)$.

For function identifiers, we define a set of auxiliary functions that retrieve relevant information about the function to make the semantics more concise: 
\begin{itemize}
\item $\Lambda_{scope} : \mathbb{IDF} -> \{global,engine,address\}$  gives the scope of the function, 
\item $\Lambda_{paraname} : \mathbb{IDF} -> {\mathbb{ID}}^{*}$ gives the function's parameter list, 
\item $\Lambda_{paratype} : \mathbb{IDF} -> {\mathbb{T}}^{*}$ gives the function's parameter types, 
\item $\Lambda_{body} : \mathbb{IDF} -> Prog$ givens the function body, and 
\item $\Lambda_{rttype} : \mathbb{IDF} -> \mathbb{T} \cup \mathit{None}$ gives the function's return type (if applicable). 
\end{itemize}

Now we introduce our notations for state modifications in the semantic rules. We use different forms of the same letter, such as with a prime, a hat or a bar, to indicate different modifications made to it. For a state $\sigma$, we use $\sigma'$ to denote the state after we apply changes to $\sigma$. Sometimes we also use $\hat{\sigma}$, $\bar{\sigma}$, $\sigma^{(1)}$, $\sigma^{(2)}$ etc to denote the intermediate state. For different letters with the same form, such as letters with primes, they represent the same modification stage. 

It is important to note that since the name space and type space of a function are auxiliary information, changes to the name space or type space will also lead to changes in the function itself.

There is a 'zoom in and out' process here. For example, if $\Psi_{i,s}$ is modified to $\Psi_{i,s}'$, then $\sigma_i$ is implicitly modified to $\sigma_i'$, with other parts remaining unchanged. This is because $\sigma_i = (\Psi_i,M_i)$ and $\Psi_i = (\Psi_{i,1},\Psi_{i,2},...,\Psi_{i,k},\Psi_{i,s})$. This convention is designed to make the formulas more concise and readable, without having to expand overly complex function definitions.

Consider the following semantic rule as an example:
\begin{gather*}
\drule{Ex}{T_{\Psi_{i,s}'}(id) = Type \\ [N_{\Psi_{i,s}'}(id)]_{\Psi_{i,s}'}^{size(Type)} = 10}{ (...,\sigma_i, \Omega_i, Prog_i ,...) -> (...,\sigma_i',\Omega_i, \cdot, ...) }
\end{gather*}
In this rule, the initial state of the $i$-th engine is $\sigma_i$ and the code to be executed is $Prog_i$, which in this context is essentially a common assignment operation. After two changes above the line, the state of the $i$-th engine becomes $\sigma_i'$ and the code to be executed becomes empty (denoted by $\cdot$). The two changes are 1) modifying the type space $T$ of the storage $\Psi_{i,s}$ for the variable $id$ to $Type$, and 2) setting the value of $id$ in the storage $\Psi_{i,s}$ to $10$. Aside from these two changes, state $\sigma_i$ and $\sigma_i'$ are the same. Additionally, the ellipsis (...) in the configuration under the line indicates that the other parts of the configuration remain unchanged.

In actual blockchain systems, all nodes in the same partition should ultimately execute the same code and have identical mempool. Therefore, in the following description, we will simplify the terminology. When we say the $i$-th engine executing certain code, it is actually the nodes in the $i$-th engine executing that code. Similarly, when we refer to the mempool of the $i$-th engine, we are actually referring to the mempools of the nodes in the $i$-th engine. Such simplification allows us to focus on the behavior at the engine level while abstracting away the details of individual node operations.

In the following we present the semantics of Crystality in rules of statements, expressions, and transactions. Common program constructs such as compositions, conditionals and loops are very similar to those in other high level languages. For space reasons, we focus on the semantics of Crystality's unique features, including variable declaration, assignment, evaluation, function calls, and relay calls. These core constructs capture the essence of Crystality's behavior, particularly its handling of different scopes and its asynchronous message passing via relay transactions.

\subsection{Semantic Rules of Statements}
The following three rules, SDa, SDs and SDg, correspond to state variable declaration statements in Crystality contracts. These statements are executed when the contract is deployed.

When a state variable is declared with the \texttt{@address} scope, an instance of the variable is created for each valid address within every engine. Appropriate storage space is allocated according to the variable's type.

In the rule SDa, the first formula indicates that initially the \texttt{@address} state variable named $id$ is not defined in $\sigma_i$. The second represents the modification of the new type space of $\sigma_i'$, assigning the type of $id$ to $Type$. The third shows that the new name space of $\sigma_i'$ is updated by allocating memory space for $id$. The last one sets the value of $id$ in the new $\sigma_i'$ to the initial value of its specified $Type$. Since each valid address has its own instance of $id$, this operation must be applied to all $k$ valid addresses within the engine.
\begin{gather*}
\drule{SDa}{ N_{\Psi_{i,j}}(id) = \mathit{None}  ,\ 1\le j\le k \\ T_{\Psi_{i,j}'}(id) = Type ,\ 1\le j\le k  \\ N_{\Psi_{i,j}'}(id) = allocate\_new(Type,\Psi_{i,j}) ,\ 1\le j\le k\\ [N_{\Psi_{i,j}'}(id)]_{\Psi_{i,j}'}^{size(Type)} = init(Type) ,\ 1\le j\le k  }{ (...,\sigma_i, \Omega_i, Type\ @address\ id; ,...) -> (...,\sigma_i',\Omega_i, \cdot, ...) }
\end{gather*}

When a state variable is declared with the \texttt{@engine} scope, an instance of the variable is created for each engine. State variable declarations are performed during contract deployment. Notably, each engine is required to execute the same statements to declare \texttt{@engine} state variables. The $i$-th engine is solely responsible for declaring its own instance of \texttt{@engine} variables.
\begin{gather*}
\drule{SDs}{ N_{\Psi_{i,s}}(id) = \mathit{None} \\ T_{\Psi_{i,s}'}(id) = Type \\\\ N_{\Psi_{i,s}'}(id) = allocate\_new(Type,\Psi_{i,s}) \\ [N_{\Psi_{i,s}'}(id)]_{\Psi_{i,s}'}^{size(Type)} = init(Type)  }{ (...,\sigma_i,\Omega_i, Type\ @engine\ id; ,...) -> (...,\sigma_i',\Omega_i,\cdot, ...) }
\end{gather*}

When a state variable is declared with the \texttt{@global} scope, an instance of the variable is created in the \texttt{@global} variable storage space $G$. The declaration of \texttt{@global} state variables involves an implicit synchronization operation. For a \texttt{@global} state variable to be successfully created in $G$, all nodes across all engines must execute the declaration. This ensures that the \texttt{@global} state variable is consistently established and maintained across all engines.

We observe hereafter that all operations involving the \texttt{@global} scope are executed consistently across all engines (with the exception of relay calls). In the configuration below the line, the code to be executed for each engine $Prog_i$ must be identical, and these $Prog_i$ must be executed simultaneously. This requirement ensures that global operations, such as reading or modifying \texttt{@global} state variables, are synchronized across the entire system.
\begin{gather*}
\drule{SDg}{N_G(id) = \mathit{None} \\ T_{G'}(id) = Type \\\\ N_{G'}(id)=allocate\_new(Type,G)\\ [N_{G'}(id)_{G'}^{size(Type)}] = init(Type) }{(\sigma_1,\Omega_1,Type\ @global\ id;,...,\sigma_n,\Omega_n,Type\ @global\ id;,G) \\ -> (\sigma_1,\Omega_1,\cdot,...,\sigma_n,\Omega_n,\cdot,G')}
\end{gather*}

The next rule TD is for temporary variable declaration. The first formula in this rule indicates that the variable is not defined within the current function. The second specifies that the newly defined variable is of type $Type$. The third one allocates storage for the new variable and the fourth initializes the new variable. The configuration of the $i$-th engine changes accordingly.
\begin{gather*}
\drule{TD}{N_{top(M_i)}(id) = \mathit{None} \\ T_{top(M_i)'}(id) = Type \\\\ N_{top(M_i)'}(id) = allocate\_new(Type,top(M_i)) \\ [N_{top(M_i)'}(id)]_{top(M_i)'}^{size(Type)} = init(Type)}{(...,\sigma_i,\Omega_i, Type\ id; ,...) -> (...,\sigma_i',\Omega_i,\cdot, ...)}
\end{gather*}

The rule TA describes the behavior of temporary variable assignment statements within a function. First, the value of the expression $exp$ is computed in the current configuration. If the corresponding temporary variable is already defined, its value in the top layer of $M$ is updated to the value of $exp$. Note that the expression $exp$ itself might also be a function call that returns a value, evaluating the value of $exp$ could lead to a change in the state of the current engine. We use a symbol with a hat $\hat{\sigma_i}$ to record this change. During the evaluation of $exp$, relay transactions may be generated, causing changes in the mempool of other engines, which is reflected by $\Omega_l$ changing to $\Omega_l'$ for every $l$.
\begin{gather*}
\drule{TA}{top(M_i).scope \ne global \\\\
 \mathcal{E}[\![(\sigma_1,\Omega_1,Prog_1,...,\sigma_i, \Omega_i, exp ,...,\sigma_n,\Omega_n,Prog_n,G)]\!] \\\\-> (\sigma_1,\Omega_1',Prog_1,...,\hat{\sigma_i},\Omega_i',v, ...,\sigma_n,\Omega_n',Prog_n,G) \\\\
N_{top(\hat{M_i})}(id) \ne \mathit{None} \;\;\;\; [N_{top(\hat{M_i}')}(id)]^{size(T_{top(\hat{M_i}')}(id))}_{top(\hat{M_i}')}=v }{(\sigma_1,\Omega_1,Prog_1,...,\sigma_i, \Omega_i, id := exp; ,...,\sigma_n,\Omega_n,Prog_n,G) \\\\-> (\sigma_1,\Omega_1',Prog_1,...,\hat{\sigma_i}',\Omega_i',\cdot, ...,\sigma_n,\Omega_n',Prog_n,G)}
\end{gather*}

Another scenario for temporary variable assignment occurs when assigning a value to a temporary variable within an \texttt{@global} function. Since \texttt{@global} functions must be executed uniformly by all engines, guaranteeing that the same operation is applied globally, a different rule is required for this case and is provided in the appendix.

We use the rule SAaa as an example for assignment statements of state variables across scopes. We need to classify the scope of the assignment statement and the scope of the involved state variables.

SAaa applies when both the function scope and the state variable scope are \texttt{@address}. In this case, the value of the expression $exp$ is computed, and the state variable is updated with the new value. Here, $\sigma_i$ and $\hat{\sigma_i}$ represent the state of the current engine before and after evaluating $exp$, respectively. Therefore, the two equations $top(...).scope$ indicate that before and after evaluating $exp$, the scopes are both address $j$.

The process involves first evaluating the expression $exp$ to obtain its value in the current configuration. The rules for evaluating expressions are presented in Section \ref{sec:rulesofevaluation}. Depending on the scope classifications, the state variable's value is then updated in its respective scope.

\begin{gather*}
\drule{SAaa}{\mathcal{E}[\![(\sigma_1,\Omega_1,Prog_1,...,\sigma_i, \Omega_i, exp ,...,\sigma_n,\Omega_n,Prog_n,G)]\!] \\\\-> (\sigma_1,\Omega_1',Prog_1,...,\hat{\sigma_i},\Omega_i',v, ...,\sigma_n,\Omega_n',Prog_n,G)  \\\\ top(M_i).scope = (address,j) \;\;\;\; top(\hat{M_i}).scope = (address,j) \\\\ N_{\hat{\Psi_{i,j}}}(id) \ne \mathit{None} \\ [N_{\hat{\Psi_{i,j}}'}(id)]^{size(T_{\hat{\Psi_{i,j}}'}(id))}_{\hat{\Psi_{i,j}}'}=v }{(\sigma_1,\Omega_1,Prog_1,...,\sigma_i, \Omega_i, id := exp; ,...,\sigma_n,\Omega_n,Prog_n,G) \\\\-> (\sigma_1,\Omega_1',Prog_1,...,\hat{\sigma_i}',\Omega_i',\cdot, ...,\sigma_n,\Omega_n',Prog_n,G)}
\end{gather*}

Due to space limitation, the rules for other assignment statements of state variables are omitted here and provided in the appendix.

Return statements can be defined through assignment statements. Each layer of $M$ is associated with a return value variable $rt$, which is a special identifier distinct from other variable names. The variable $rt$ is declared immediately when the function is called, and its type is the return type of the function. Thus, the return value can be accessed through $rt$. The rules corresponding to functions with return values are presented in Section \ref{sec:rulesofevaluation}.

In the following, we use the rule IFaa as an example for function calls without return values. IFaa describes the scenario where a function with scope \texttt{@address} calls another function also with scope \texttt{@address}. During the execution of the function body, relay transactions may be generated, causing changes in the mempool of other engines, which is reflected by $\Omega_l$ changing to $\Omega_l'$ for every $l$.
\begin{gather*}
\drule{IFaa}{\Lambda_{scope}(idf)=address \\ top(M_i).scope = (address,j)\\\\
\Lambda_{rettype}(idf) = \mathit{None} \\ \Lambda_{body}(idf) = Block \\\\
\Lambda_{paraname}(idf) =(id_1,...,id_t) \\ \Lambda_{paratype}(idf) = (T_1,...,T_t)\\\\
M_i^{(1)}=push(M_i,top(M_i)) \\ top(M_i^{(1)}).scope = (address,j) \\\\
top(M_i^{(1)}).rt = \mathit{None} \\ M_i^{(3)} = pop(M_i^{(2)})\\\\
(\sigma_1,\Omega_1,Prog_1,...,\sigma_i^{(1)},\Omega_i,T_1\ id_1 := exp_1;\\...;T_t\ id_t := exp_t;Block,...,\sigma_n,\Omega_n,Prog_n,G) \\-> (\sigma_1,\Omega_1',Prog_1,...,\sigma_i^{(2)},\Omega_i',\cdot,...,\sigma_n,\Omega_n',Prog_n,G) 
}{(\sigma_1,\Omega_1,Prog_1,...,\sigma_i,\Omega_i,idf(exp_1,...,exp_t),...,\sigma_n,\Omega_n,Prog_n,G) \\-> (\sigma_1,\Omega_1',Prog_1,...,\sigma_i^{(3)},\Omega_i',\cdot,...,\sigma_n,\Omega_n',Prog_n,G)}
\end{gather*}

\texttt{@global} functions are executed on all nodes across all engines, ensuring that \texttt{@global} state variables remain consistent throughout the entire system. When a \texttt{@global} function is called, each node in every engine executes the function independently, and the result is a synchronized and uniform update to the \texttt{@global} state variables. The storage $G$ may change, but the configuration of each engine cannot change because \texttt{@global} function is not allowed to modify \texttt{@engine} variables or \texttt{@address} variables of each engine.

For brevity, the formal rules for other function calls are omitted here but are fully detailed in the appendix.

The following rules (from RELa to RELg1) describe the behavior of relay calls. We formalize the rules for the packaging and propagation of the relay transactions.

A relay call is akin to a function call, but it is asynchronous. The call data is packaged into a relay transaction. The relay statement itself returns immediately.  There are three types of relay targets, determining how the relay transaction is processed and executed on the target engine:

RELa specifies a specific target address. 
The evaluation of $exp$ represents the corresponding address location. The result of the evaluation is $(r, j)$, which denotes the $j$-th address of the $r$-th engine. The function $idf$ with the given arguments is packaged into a relay transaction and sent to the engine responsible for the specific target address.
\begin{gather*}
\drule{RELa}{\Lambda_{scope}(idf) = address \\ \mathcal{E}[\![(...,\sigma_i,\Omega_i,exp,...)]\!] -> (...,\sigma_i,\Omega_i,(r,j),...) \\
\mathcal{E}[\![(...,\sigma_i,\Omega_i,exp_l,...)]\!] -> (...,\sigma_i,\Omega_i,v_l,...),\ 1 \le l \le t \\
\Omega_r'=\Omega_r \cup \{ (address,j,idf(v_1,...,v_t)) \} 
}{(...,\sigma_i,\Omega_i,relay @exp\ idf(exp_1,...,exp_t);,...,\sigma_r,\Omega_r,Prog_r...) \\\\ -> (...,\sigma_i,\Omega_i,\cdot,...,\sigma_r,\Omega_r',Prog_r,...)}
\end{gather*}

RELs acts like broadcast, relaying from one engine to all engines. Here the called function must have scope \texttt{@engine}. This statement adds a relay transaction to the mempool of every engine. 
\begin{gather*}
\drule{RELs}{\Lambda_{scope}(idf) = engine \\\\
\mathcal{E}[\![(...,\sigma_i,\Omega_i,exp_l,...)]\!] -> (...,\sigma_i,\Omega_i,v_l,...),\ 1 \le l \le t\\\\
\Omega_i'=\Omega_i \cup \{ (engine,idf(v_1,...,v_t)) \} ,\ 1 \le i \le n}{(\sigma_1,\Omega_1,Prog_1,...,\sigma_i,\Omega_i,relay @engines\ idf(exp_1,...,exp_t);,\\...,\sigma_n,\Omega_n,Prog_n,G)\\\\->(\sigma_1,\Omega_1',Prog_1,...,\sigma_i,\Omega_i',\cdot,...,\sigma_n,\Omega_n',Prog_n,G)}
\end{gather*}

The next rule also involves relaying to all engines, but it differ slightly from the previous one. Firstly, the functions invoked in this rules are \texttt{@global} functions. Secondly, the relay statements must be called from functions that have an \texttt{@address} or \texttt{@engine} scope, i.e. they cannot be invoked from a \texttt{@global} function. This restriction ensures that \texttt{@global} functions are only triggered by relay transactions initiated from more localized scopes.
\begin{gather*}
\drule{RELg1}{\Lambda_{scope}(idf) = global \\ top(M_i).scope = (address,j) \\\\
\mathcal{E}[\![(...,\sigma_i,\Omega_i,exp_l,...)]\!] -> (...,\sigma_i,\Omega_i,v_l,...),\ 1 \le l \le t\\\\
\Omega_i'=\Omega_i \cup \{ (global,idf(v_1,...,v_t)) \} ,\ 1 \le i \le n}{(\sigma_1,\Omega_1,Prog_1,...,\sigma_i,\Omega_i,relay @global\ idf(exp_1,...,exp_t);,\\...,\sigma_n,\Omega_n,Prog_n,G) \\\\->(\sigma_1,\Omega_1',Prog_1,...,\sigma_i,\Omega_i',\cdot,...,\sigma_n,\Omega_n',Prog_n,G)}
\end{gather*}

We formalize the execution rules governing Crystality's parallel execution paradigm. Specifically, the Para rule captures the parallel execution across multiple execution engines, allowing concurrent state updates in different engines. Additionally, we introduce the MemP rule to define mempool behavior in Crystality. This rule ensures that relay transactions in the mempool do not affect the execution of the current transaction, maintaining execution isolation from the mempool of each engine. These rules can be found in the appendix.

\subsection{Semantic Rules of Evaluations} \label{sec:rulesofevaluation}
The semantics of accessing and reading of temporary variables is captured by the following rule TE. In any engine, the value of a temporary variable within the current function can be read as long as the variable is defined.
\begin{gather*}
\drule{TE}{N_{top(M_i)}(id) \ne \mathit{None} \\
[N_{top(M_i)}(id)]^{size(T_{top(M_i)}(id))}_{top(M_i)}=v }{\mathcal{E}[\![(...,\sigma_i,\Omega_i,id,...) ]\!]->(...,\sigma_i,\Omega_i,v,...)}
\end{gather*}

The following rules are related to reading state variables. Access to state variables must be determined based on the scope of the state variable and the scope of the current function. The permissions for functions to read state variables vary depending on their scope. For further details refer to Table~\ref{tab1}. Here, we provide two rules, SEaa and SEgg, as examples for illustration. Other rules are included in the appendix.

At the $j$-th address in the $i$-th engine, the function can read the \texttt{@address} state variable of the $j$-th address in the $i$-th engine. 
\begin{gather*}
\drule{SEaa}{top(M_i).scope = (address,j) \\ N_{\Psi_{i,j}}(id) \ne \mathit{None}\\
[N_{\Psi_{i,j}}(id)]^{size(T_{\Psi_{i,j}}(id))}_{\Psi_{i,j}}=v }{\mathcal{E}[\![(...,\sigma_i,\Omega_i,id,...) ]\!]->(...,\sigma_i,\Omega_i,v,...)}
\end{gather*}

The evaluation within \texttt{@global} functions is performed on all nodes across all engines. 
\begin{gather*}
\drule{SEgg}{top(M_i).scope = global \\ N_G(id) \ne \mathit{None}\\
[N_G(id)]^{size(T_G(id))}_G=v }{\mathcal{E}[\![(\sigma_1,\Omega_1,id,...,\sigma_n,\Omega_n,id,G) ]\!] ->(\sigma_1,\Omega_1,v,...,\sigma_n,\Omega_n,v,G)}
\end{gather*}

The rules about function calls with return values are similar to the previous rules for function calls without return values, but require the use of variable $rt$ in each layer of $M$ to record the return value. This ensures that the return value is captured properly and can be accessed after the function execution. Under normal circumstances, the return value of calls to \texttt{@global} functions should be the same across all engines. This consistency is ensured by the consensus mechanism. Additionally, since \texttt{@global} functions can only modify \texttt{@global} state variables, the configuration of each engine remains unchanged before and after the function call. However, because the function may contain relay statements, the mempool of each engine may change. Owing to space limitations, we present the rules for function calls with return values in the appendix.

\subsection{Semantic Rules of Transactions}
Defining the semantics of transactions is fundamental to smart contract languages. We model a transaction as a function call, abstracting details such as gas and signatures while focusing on its functional behavior. To formalize this, we introduce T-functions, which encapsulate the essential execution logic of transactions. In the following, we outline the constraints that T-functions must satisfy and provide their precise mathematical definition.

Since a transaction is the outermost function, the current scope of the top layer of $M$ must be $\mathit{None}$. The nodes can get the address of the sender of this transaction using function $get\_sender\_address()$ from the information attached to the transaction, and determine which engine the address is in. Then a new layer is built on top of stack $M$ to denote the invocation of function, and records the scope. Upon function return, this layer is removed. We also need to record the changes of mempools of other engines because there may be relay statements in function body.

Crystality supports both user-initiated and relay transactions. The originating engine of a relay transaction is also identified via $get\_sender\_address()$, enabling a uniform handling mechanism for both types.

The impact of a transaction on the world state is reflected in each engine's storage $\Psi_i$, the storage space of \texttt{@global} state variables $G$, and each engine's mempool $\Omega_i$, but not in each engine's memory $M_i$. This is because memory holds the values of temporary variables specific to the engine, which are created during function execution and deleted upon function return. Since the T-function invoked by the transaction is the outermost function, the memory stack of each engine is cleared when the transaction completes, ensuring memory is empty both before and after transaction execution.

The formal semantic rules are detailed in the appendix. For instance, the IFta rule specifies the behavior when a transaction invokes a function with the scope \texttt{@address} within a contract.

\section{Formalization and Verification of Crystality in Coq}
Coq \cite{huet1997coq} is an interactive theorem prover based on higher-order logic, widely used for formal verification in programming languages, mathematics, and critical software systems. Unlike HOL4 \cite{slind2008brief}, Isabelle/HOL \cite{foster2015isabelle}, and PVS \cite{owre1992pvs}, Coq is built on the Calculus of Inductive Constructions, integrating higher-order logic with a richly typed functional programming language. It supports structural proof scripting and proof automation, making it effective for verifying properties like type soundness, program correctness, and compiler certification. Notable applications include CompCert \cite{krebbers2014formal}, JavaCard security verification \cite{andronick2003using}, and the Bedrock framework \cite{chlipala2013bedrock}. For a detailed introduction, see \cite{bertot2013interactive}.

We formalize Crystality’s core syntax and semantics in Coq and verify a simplified ERC20 token transfer function. Specifically, we prove that if the transfer amount does not exceed the sender’s balance, the sender’s balance decreases accordingly, and a relay transaction is created in the recipient’s engine to increase the recipient’s balance.
If the transfer amount exceeds the sender’s balance, the transaction fails, leaving the ledger state unchanged. Furthermore, using our semantic rules, we verify the refinement relationship between two token transfer functions. Due to space constraints, the full Coq formalization and verification are not provided here and available at \cite{crystality_coq}.

While our current Coq formalization faithfully captures the core operational semantics of Crystality, it remains a foundational model with several limitations. In future work, we plan to significantly enrich the formalization to account for the full feature set of Crystality, particularly the semantics of parallelism and synchronization. We also intend to evaluate our semantics against more complex, real-world contracts to validate the expressiveness and applicability of our model.

\section{Related Work}
Although formal semantics for smart contract programming languages has garnered some attention, most of the existing research has concentrated on languages with sequential execution models, particularly Solidity. This includes a denotational semantics of Solidity proposed in \cite{Marmsoler}, a big-step semantics of a subset of Solidity formalized in \cite{zakrzewski2018towards}, an operational semantics for a subset of Solidity described in \cite{crosaratowards}, an executable operational semantics of Solidity presented in \cite{jiao2018,jiao2020semantic}, a formalization of Solidity in terms of a SMT-based intermediate language provided in \cite{Hajdu1,Hajdu2}, and a formal semantics for an intermediate specification language for the formal verification of Ethereum-based smart contracts in Coq, proposed in \cite{Yang2020}. Additionally, \cite{hildenbrandt2018kevm} provides formal semantics for EVM using the K-framework. None of these are directly applicable to smart contract languages designed for parallel EVMs.

\section{Conclusion}
This paper presents the structural operational semantics for Crystality, a novel smart contract programming language designed for parallel EVMs. Our work provides a formal execution model that precisely captures Crystality’s key features, including Programmable Contract Scopes and Asynchronous Functional Relay, which enable fine-grained state partitioning and efficient inter-scope communication. By formalizing the core syntax and semantics of Crystality in Coq, we establish a rigorous foundation for mechanized reasoning and verification of smart contracts. The verification of a simple token transfer function demonstrates the applicability of our framework in ensuring smart contract correctness.

Several avenues remain for future exploration. Extending the semantics to incorporate gas models and execution costs would enhance practical applicability. Strengthening security verification by formally proving protection against reentrancy attacks, front-running vulnerabilities, and state consistency violations in parallel execution settings is another key avenue. Additionally, establishing semantic correspondence with languages like Solidity would improve interoperability. Finally, developing a verified compiler that translates high-level Crystality code into optimized EVM bytecode would bridge the gap between formal semantics and real-world execution.

Addressing these directions will advance the formalization, verification, and practical adoption of parallel smart contract languages, contributing to scalable and high-assurance blockchain computing.

\begin{credits}
\subsubsection{\ackname} This work was supported by the National Key R\&D Program of China under Grant 2022YFB2702200, and the NSFC under Grant 62172019.
\end{credits}

\bibliographystyle{splncs04}
\bibliography{mypaper}

\appendix

\section{Semantic Rules of Statements}

\begin{gather*}
\drule{MemP}{\Omega_i' \setminus \Omega_i  = \bar{\Omega_i} \setminus \hat{\Omega_i} \\\\ (\sigma_1,\Omega_1,Prog_1,...,\sigma_i,\Omega_i,Prog_i,...,\sigma_n,\Omega_n,Prog_n,G) \\\\-> (\sigma_1',\Omega_1',Prog_1',...,\sigma_i',\Omega_i',Prog_i',...,\sigma_n',\Omega_n',Prog_n',G')}{(\sigma_1,\Omega_1,Prog_1,...,\sigma_i,\hat{\Omega_i},Prog_i,...,\sigma_n,\Omega_n,Prog_n,G) \\\\-> (\sigma_1',\Omega_1',Prog_1',,...,\sigma_i',\bar{\Omega_i},Prog_i',...,\sigma_n',\Omega_n',Prog_n',G')} 
\end{gather*}

\begin{gather*}
\drule{Para}{\bar{\Omega_r} \cup \hat{\Omega_r} = \Omega_r',\ 1 \le r \le n\\\\(\sigma_1,\Omega_1,Prog_1,...,\sigma_i,\Omega_i,Prog_i,...,\sigma_j,\Omega_j,Prog_j,...,\sigma_n,\Omega_n,Prog_n,G) \\\\-> (\sigma_1,\bar{\Omega_1},Prog_1,...,\sigma_i',\bar{\Omega_i},\cdot,...,\sigma_j,\bar{\Omega_j},Prog_j,...,\sigma_n,\bar{\Omega_n},Prog_n,G)\\\\(\sigma_1,\Omega_1,Prog_1,...,\sigma_i,\Omega_i,Prog_i,...,\sigma_j,\Omega_j,Prog_j,...,\sigma_n,\Omega_n,Prog_n,G) \\\\-> (\sigma_1,\hat{\Omega_1},Prog_1,...,\sigma_i,\hat{\Omega_i},Prog_i,...,\sigma_j',\hat{\Omega_j},\cdot,...,\sigma_n,\hat{\Omega_n},Prog_n,G)}{(\sigma_1,\Omega_1,Prog_1,...,\sigma_i,\Omega_i,Prog_i,...,\sigma_j,\Omega_j,Prog_j,...,\sigma_n,\Omega_n,Prog_n,G) \\\\->(\sigma_1,\Omega_1',Prog_1,...,\sigma_i',\Omega_i',\cdot,...,\sigma_j',\Omega_j',\cdot,...,\sigma_n,\Omega_n',Prog_n,G)}
\end{gather*}

\begin{gather*}
\drule{SDa}{ N_{\Psi_{i,j}}(id) = \mathit{None}  ,\ 1\le j\le k \\ T_{\Psi_{i,j}'}(id) = Type ,\ 1\le j\le k  \\ N_{\Psi_{i,j}'}(id) = allocate\_new(Type,\Psi_{i,j}) ,\ 1\le j\le k\\ [N_{\Psi_{i,j}'}(id)]_{\Psi_{i,j}'}^{size(Type)} = init(Type) ,\ 1\le j\le k  }{ (...,\sigma_i, \Omega_i, Type\ @address\ id; ,...) -> (...,\sigma_i',\Omega_i, \cdot, ...) }
\end{gather*}

\begin{gather*}
\drule{SDs}{ N_{\Psi_{i,s}}(id) = \mathit{None} \\ T_{\Psi_{i,s}'}(id) = Type \\\\ N_{\Psi_{i,s}'}(id) = allocate\_new(Type,\Psi_{i,s}) \\ [N_{\Psi_{i,s}'}(id)]_{\Psi_{i,s}'}^{size(Type)} = init(Type)  }{ (...,\sigma_i,\Omega_i, Type\ @engine\ id; ,...) -> (...,\sigma_i',\Omega_i,\cdot, ...) }
\end{gather*}

\begin{gather*}
\drule{SDg}{N_G(id) = \mathit{None} \\ T_{G'}(id) = Type \\\\ N_{G'}(id)=allocate\_new(Type,G)\\ [N_{G'}(id)_{G'}^{size(Type)}] = init(Type) }{(\sigma_1,\Omega_1,Type\ @global\ id;,...,\sigma_n,\Omega_n,Type\ @global\ id;,G) \\ -> (\sigma_1,\Omega_1,\cdot,...,\sigma_n,\Omega_n,\cdot,G')}
\end{gather*}

\begin{gather*}
\drule{TD}{N_{top(M_i)}(id) = \mathit{None} \\ T_{top(M_i)'}(id) = Type \\\\ N_{top(M_i)'}(id) = allocate\_new(Type,top(M_i)) \\ [N_{top(M_i)'}(id)]_{top(M_i)'}^{size(Type)} = init(Type)}{(...,\sigma_i,\Omega_i, Type\ id; ,...) -> (...,\sigma_i',\Omega_i,\cdot, ...)}
\end{gather*}

\begin{gather*}
\drule{TA}{top(M_i).scope \ne global \\\\
 \mathcal{E}[\![(\sigma_1,\Omega_1,Prog_1,...,\sigma_i, \Omega_i, exp ,...,\sigma_n,\Omega_n,Prog_n,G)]\!] \\\\-> (\sigma_1,\Omega_1',Prog_1,...,\hat{\sigma_i},\Omega_i',v, ...,\sigma_n,\Omega_n',Prog_n,G) \\\\
N_{top(\hat{M_i})}(id) \ne \mathit{None} \;\;\;\; [N_{top(\hat{M_i}')}(id)]^{size(T_{top(\hat{M_i}')}(id))}_{top(\hat{M_i}')}=v }{(\sigma_1,\Omega_1,Prog_1,...,\sigma_i, \Omega_i, id := exp; ,...,\sigma_n,\Omega_n,Prog_n,G) \\\\-> (\sigma_1,\Omega_1',Prog_1,...,\hat{\sigma_i}',\Omega_i',\cdot, ...,\sigma_n,\Omega_n',Prog_n,G)}
\end{gather*}

\begin{gather*}
\drule{TAg}{top(M_i).scope = global ,\ 1 \le i \le n  \\ \mathcal{E}[\![(\sigma_1,\Omega_1,exp,...,\sigma_n,\Omega_n,exp,G)]\!] -> (\sigma_1,\Omega_1',v,...,\sigma_n,\Omega_n',v,G') \\\\
N_{top(M_i)}(id) \ne \mathit{None} ,\ 1 \le i \le n  \\ [N_{top(M_i')}(id)]^{size(T_{top(M_i')}(id))}_{top(M_i')}=v ,\ 1 \le i \le n }{(\sigma_1,\Omega_1,id := exp; ,...,\sigma_n,\Omega_n,id := exp; ,G) -> (\sigma_1',\Omega_1',\cdot,...,\sigma_n',\Omega_n',\cdot,G')}
\end{gather*}

\begin{gather*}
\drule{SAaa}{\mathcal{E}[\![(\sigma_1,\Omega_1,Prog_1,...,\sigma_i, \Omega_i, exp ,...,\sigma_n,\Omega_n,Prog_n,G)]\!] \\\\-> (\sigma_1,\Omega_1',Prog_1,...,\hat{\sigma_i},\Omega_i',v, ...,\sigma_n,\Omega_n',Prog_n,G)  \\\\ top(M_i).scope = (address,j) \;\;\;\; top(\hat{M_i}).scope = (address,j) \\\\ N_{\hat{\Psi_{i,j}}}(id) \ne \mathit{None} \\ [N_{\hat{\Psi_{i,j}}'}(id)]^{size(T_{\hat{\Psi_{i,j}}'}(id))}_{\hat{\Psi_{i,j}}'}=v }{(\sigma_1,\Omega_1,Prog_1,...,\sigma_i, \Omega_i, id := exp; ,...,\sigma_n,\Omega_n,Prog_n,G) \\\\-> (\sigma_1,\Omega_1',Prog_1,...,\hat{\sigma_i}',\Omega_i',\cdot, ...,\sigma_n,\Omega_n',Prog_n,G)}
\end{gather*}

\begin{gather*}
\drule{SAas}{\mathcal{E}[\![(\sigma_1,\Omega_1,Prog_1,...,\sigma_i, \Omega_i, exp ,...,\sigma_n,\Omega_n,Prog_n,G)]\!] \\\\-> (\sigma_1,\Omega_1',Prog_1,...,\hat{\sigma_i},\Omega_i',v, ...,\sigma_n,\Omega_n',Prog_n,G)  \\\\  top(M_i).scope = (address,j) \;\;\;\; top(\hat{M_i}).scope = (address,j) \\\\ N_{\hat{\Psi_{i,s}}}(id) \ne \mathit{None} \\ [N_{\hat{\Psi_{i,s}}'}(id)]^{size(T_{\hat{\Psi_{i,s}}'}(id))}_{\hat{\Psi_{i,s}}'}=v }{(\sigma_1,\Omega_1,Prog_1,...,\sigma_i, \Omega_i, id := exp; ,...,\sigma_n,\Omega_n,Prog_n,G) \\\\-> (\sigma_1,\Omega_1',Prog_1,...,\hat{\sigma_i}',\Omega_i',\cdot, ...,\sigma_n,\Omega_n',Prog_n,G)}
\end{gather*}

\begin{gather*}
\drule{SAss}{\mathcal{E}[\![(\sigma_1,\Omega_1,Prog_1,...,\sigma_i, \Omega_i, exp ,...,\sigma_n,\Omega_n,Prog_n,G)]\!] \\\\-> (\sigma_1,\Omega_1',Prog_1,...,\hat{\sigma_i},\Omega_i',v, ...,\sigma_n,\Omega_n',Prog_n,G)  \\ top(M_i).scope = engine \\ top(\hat{M_i}).scope = engine \\\\  N_{\hat{\Psi_{i,s}}}(id) \ne \mathit{None} \\ [N_{\hat{\Psi_{i,s}}'}(id)]^{size(T_{\hat{\Psi_{i,s}}'}(id))}_{\hat{\Psi_{i,s}}'}=v }{(\sigma_1,\Omega_1,Prog_1,...,\sigma_i, \Omega_i, id := exp; ,...,\sigma_n,\Omega_n,Prog_n,G) \\\\-> (\sigma_1,\Omega_1',Prog_1,...,\hat{\sigma_i}',\Omega_i',\cdot, ...,\sigma_n,\Omega_n',Prog_n,G)}
\end{gather*}

\begin{gather*}
\drule{SAgg}{\mathcal{E}[\![(\sigma_1,\Omega_1,exp,...,\sigma_n,\Omega_n,exp,G)]\!] -> (\sigma_1,\Omega_1',v,...,\sigma_n,\Omega_n',v,\hat{G})  \\ top(M_i).scope = global,\ 1 \le i \le n \\  N_{\hat{G}}(id) \ne \mathit{None} \\ [N_{\hat{G}'}(id)]^{size(T_{\hat{G}'}(id))}_{\hat{G}'}=v }{(\sigma_1,\Omega_1,id := exp;,...,\sigma_n,\Omega_n,id := exp;,G) -> (\sigma_1,\Omega_1',\cdot,...,\sigma_n,\Omega_n',\cdot,\hat{G}')}
\end{gather*}

\begin{gather*}
\drule{RET}{top(M).rt = rt \\\\ (\sigma_1,\Omega_1,Prog_1,...,\sigma_i, \Omega_i, rt := exp; ,...,\sigma_n,\Omega_n,Prog_n,G)\\ -> (\sigma_1,\Omega_1',Prog_1,...,\sigma_i',\Omega_i',\cdot, ...,\sigma_n,\Omega_n',Prog_n,G)}{(\sigma_1,\Omega_1,Prog_1,...,\sigma_i, \Omega_i, return\ exp; ,...,\sigma_n,\Omega_n,Prog_n,G) \\\\-> (\sigma_1,\Omega_1',Prog_1,...,\sigma_i',\Omega_i',\cdot, ...,\sigma_n,\Omega_n',Prog_n,G) }
\end{gather*}

\begin{gather*}
\drule{IFaa}{\Lambda_{scope}(idf)=address \\ top(M_i).scope = (address,j)\\\\
\Lambda_{rettype}(idf) = \mathit{None} \\ \Lambda_{body}(idf) = Block \\\\
\Lambda_{paraname}(idf) =(id_1,...,id_t) \\ \Lambda_{paratype}(idf) = (T_1,...,T_t)\\\\
M_i^{(1)}=push(M_i,top(M_i)) \\ top(M_i^{(1)}).scope = (address,j) \\\\
top(M_i^{(1)}).rt = \mathit{None} \\ M_i^{(3)} = pop(M_i^{(2)})\\\\
(\sigma_1,\Omega_1,Prog_1,...,\sigma_i^{(1)},\Omega_i,T_1\ id_1 := exp_1;\\...;T_t\ id_t := exp_t;Block,...,\sigma_n,\Omega_n,Prog_n,G) \\-> (\sigma_1,\Omega_1',Prog_1,...,\sigma_i^{(2)},\Omega_i',\cdot,...,\sigma_n,\Omega_n',Prog_n,G) 
}{(\sigma_1,\Omega_1,Prog_1,...,\sigma_i,\Omega_i,idf(exp_1,...,exp_t),...,\sigma_n,\Omega_n,Prog_n,G) \\-> (\sigma_1,\Omega_1',Prog_1,...,\sigma_i^{(3)},\Omega_i',\cdot,...,\sigma_n,\Omega_n',Prog_n,G)}
\end{gather*}

\begin{gather*}
\drule{IFas}{\Lambda_{scope}(idf)=engine \\ top(M_i).scope = (address,j)\\\\
\Lambda_{rettype}(idf) = \mathit{None} \\ \Lambda_{body}(idf) = Block \\\\
\Lambda_{paraname}(idf) =(id_1,...,id_t) \\ \Lambda_{paratype}(idf) = (T_1,...,T_t)\\\\ 
M_i^{(1)}=push(M_i,top(M_i)) \\ top(M_i^{(1)}).scope = engine \\\\
top(M_i^{(1)}).rt = \mathit{None} \\ M_i^{(3)} = pop(M_i^{(2)})\\\\
(\sigma_1,\Omega_1,Prog_1,...,\sigma_i^{(1)},\Omega_i,T_1\ id_1 := exp_1;\\...;T_t\ id_t := exp_t;Block,...,\sigma_n,\Omega_n,Prog_n,G) \\ -> (\sigma_1,\Omega_1',Prog_1,...,\sigma_i^{(2)},\Omega_i',\cdot,...,\sigma_n,\Omega_n',Prog_n,G) 
}{(\sigma_1,\Omega_1,Prog_1,...,\sigma_i,\Omega_i,idf(exp_1,...,exp_t),...,\sigma_n,\Omega_n,Prog_n,G) \\\\-> (\sigma_1,\Omega_1',Prog_1,...,\sigma_i^{(3)},\Omega_i',\cdot,...,\sigma_n,\Omega_n',Prog_n,G)}
\end{gather*}

\begin{gather*}
\drule{IFss}{\Lambda_{scope}(idf)=engine \\ top(M_i).scope = engine\\\\
\Lambda_{rettype}(idf) = \mathit{None} \\ \Lambda_{body}(idf) = Block \\\\
\Lambda_{paraname}(idf) =(id_1,...,id_t) \\ \Lambda_{paratype}(idf) = (T_1,...,T_t)\\\\
M_i^{(1)}=push(M_i,top(M_i)) \\ top(M_i^{(1)}).scope = engine \\\\
top(M_i^{(1)}).rt = \mathit{None} \\ M_i^{(3)} = pop(M_i^{(2)})\\\\
(\sigma_1,\Omega_1,Prog_1,...,\sigma_i^{(1)},\Omega_i,T_1\ id_1 := exp_1;\\...;T_t\ id_t := exp_t;Block,...,\sigma_n,\Omega_n,Prog_n,G) \\-> (\sigma_1,\Omega_1',Prog_1,...,\sigma_i^{(2)},\Omega_i',\cdot,...,\sigma_n,\Omega_n',Prog_n,G) 
}{(\sigma_1,\Omega_1,Prog_1,...,\sigma_i,\Omega_i,idf(exp_1,...,exp_t),...,\sigma_n,\Omega_n,Prog_n,G) \\\\-> (\sigma_1,\Omega_1',Prog_1,...,\sigma_i^{(3)},\Omega_i',\cdot,...,\sigma_n,\Omega_n',Prog_n,G)}
\end{gather*}

\begin{gather*}
\drule{IFgg}{\Lambda_{scope}(idf)=global \\ top(M_i).scope = global\\\\
\Lambda_{rettype}(idf) = \mathit{None} \\ \Lambda_{body}(idf) = Block \\\\
\Lambda_{paraname}(idf) =(id_1,...,id_t) \\ \Lambda_{paratype}(idf) = (T_1,...,T_t)\\\\
M_i^{(1)}=push(M_i,top(M_i)),\ 1 \le i \le n \\\\
top(M_i^{(1)}).scope = global  ,\ 1 \le i \le n \\
top(M_i^{(1)}).rt = \mathit{None} ,\ 1 \le i \le n \\\\
M_i^{(3)} = pop(M_i^{(2)}) ,\ 1 \le i \le n \\
\sigma_i^{(3)} = \sigma_i ,\ 1 \le i \le n\\\\
(\sigma_1^{(1)},\Omega_1,T_1\ id_1 := exp_1;...;T_t\ id_t := exp_t;Block,...,\\\\\sigma_n^{(1)},\Omega_n,T_1\ id_1 := exp_1;...;T_t\ id_t := exp_t;Block,G) \\\\-> (\sigma_1^{(2)},\Omega_1',\cdot,...,\sigma_n^{(2)},\Omega_n',\cdot,G') }{(\sigma_1,\Omega_1,idf(exp_1,...,exp_t),...,\sigma_n,\Omega_n,idf(exp_1,...,exp_t),G) \\ -> (\sigma_1,\Omega_1',\cdot,...,\sigma_n,\Omega_n',\cdot,G')}
\end{gather*}

\begin{gather*}
\drule{RELa}{\Lambda_{scope}(idf) = address \\ \mathcal{E}[\![(...,\sigma_i,\Omega_i,exp,...)]\!] -> (...,\sigma_i,\Omega_i,(r,j),...) \\
\mathcal{E}[\![(...,\sigma_i,\Omega_i,exp_l,...)]\!] -> (...,\sigma_i,\Omega_i,v_l,...),\ 1 \le l \le t \\
\Omega_r'=\Omega_r \cup \{ (address,j,idf(v_1,...,v_t)) \} 
}{(...,\sigma_i,\Omega_i,relay @exp\ idf(exp_1,...,exp_t);,...,\sigma_r,\Omega_r,Prog_r...) \\\\ -> (...,\sigma_i,\Omega_i,\cdot,...,\sigma_r,\Omega_r',Prog_r,...)}
\end{gather*}

\begin{gather*}
\drule{RELs}{\Lambda_{scope}(idf) = engine \\\\
\mathcal{E}[\![(...,\sigma_i,\Omega_i,exp_l,...)]\!] -> (...,\sigma_i,\Omega_i,v_l,...),\ 1 \le l \le t\\\\
\Omega_i'=\Omega_i \cup \{ (engine,idf(v_1,...,v_t)) \} ,\ 1 \le i \le n}{(\sigma_1,\Omega_1,Prog_1,...,\sigma_i,\Omega_i,relay @engines\ idf(exp_1,...,exp_t);,\\...,\sigma_n,\Omega_n,Prog_n,G)\\\\->(\sigma_1,\Omega_1',Prog_1,...,\sigma_i,\Omega_i',\cdot,...,\sigma_n,\Omega_n',Prog_n,G)}
\end{gather*}

\begin{gather*}
\drule{RELg1}{\Lambda_{scope}(idf) = global \\ top(M_i).scope = (address,j) \\\\
\mathcal{E}[\![(...,\sigma_i,\Omega_i,exp_l,...)]\!] -> (...,\sigma_i,\Omega_i,v_l,...),\ 1 \le l \le t\\\\
\Omega_i'=\Omega_i \cup \{ (global,idf(v_1,...,v_t)) \} ,\ 1 \le i \le n}{(\sigma_1,\Omega_1,Prog_1,...,\sigma_i,\Omega_i,relay @global\ idf(exp_1,...,exp_t);,\\...,\sigma_n,\Omega_n,Prog_n,G) \\\\->(\sigma_1,\Omega_1',Prog_1,...,\sigma_i,\Omega_i',\cdot,...,\sigma_n,\Omega_n',Prog_n,G)}
\end{gather*}

\begin{gather*}
\drule{RELg2}{\Lambda_{scope}(idf) = global \\ top(M_i).scope = engine \\\\
\mathcal{E}[\![(...,\sigma_i,\Omega_i,exp_l,...)]\!] -> (...,\sigma_i,\Omega_i,v_l,...),\ 1 \le l \le t\\\\
\Omega_i'=\Omega_i \cup \{ (global,idf(v_1,...,v_t)) \} ,\ 1 \le i \le n}{(\sigma_1,\Omega_1,Prog_1,...,\sigma_i,\Omega_i,relay@global\ idf(exp_1,...,exp_t);,\\...,\sigma_n,\Omega_n,Prog_n,G) \\\\->(\sigma_1,\Omega_1',Prog_1,...,\sigma_i,\Omega_i',\cdot,...,\sigma_n,\Omega_n',Prog_n,G)}
\end{gather*}

\begin{gather*}
\drule{SEQ}{(\sigma_1,\Omega_1,Prog_1,...,\sigma_i,\Omega_i,Stmt1;,...,\sigma_n,\Omega_n,Prog_n,G)\\\\->(\sigma_1,\hat{\Omega_1},Prog_1,...,\hat{\sigma_i},\hat{\Omega_i},\cdot,...,\sigma_n,\hat{\Omega_n},Prog_n,G)\\ (\sigma_1,\hat{\Omega_1},Prog_1,...,\hat{\sigma_i},\hat{\Omega_i},Stmt2;,...,\sigma_n,\hat{\Omega_n},Prog_n,G) \\\\-> (\sigma_1,\hat{\Omega_1}',Prog_1,...,\hat{\sigma_i}',\hat{\Omega_i}',\cdot,...,\sigma_n,\hat{\Omega_n}',Prog_n,G)  }{(\sigma_1,\Omega_1,Prog_1,...,\sigma_i,\Omega_i,Stmt1;Stmt2;,...,\sigma_n,\Omega_n,Prog_n,G)\\\\->(\sigma_1,\hat{\Omega_1}',Prog_1,...,\hat{\sigma_i}',\hat{\Omega_i}',\cdot,...,\sigma_n,\hat{\Omega_n}',Prog_n,G)}
\end{gather*}

\begin{gather*}
\drule{COND1}{ \mathcal{E}[\![ (\sigma_1,\Omega_1,Prog_1,...,\sigma_i,\Omega_i,exp,...,\sigma_n,\Omega_n,Prog_n,G) ]\!] \\\\-> (\sigma_1,\hat{\Omega_1},Prog_1,...,\hat{\sigma_i},\hat{\Omega_i},\mathbf{True},...,\sigma_n,\hat{\Omega_n},Prog_n,G)\\ (\sigma_1,\hat{\Omega_1},Prog_1,...,\hat{\sigma_i},\hat{\Omega_i},Block_1,...,\sigma_n,\hat{\Omega_n},Prog_n,G) \\\\-> (\sigma_1,\hat{\Omega_1}',Prog_1,...,\hat{\sigma_i}',\hat{\Omega_i}',\cdot,...,\sigma_n,\hat{\Omega_n}',Prog_n,G) }{(\sigma_1,\Omega_1,Prog_1,...,\sigma_i,\Omega_i,if\ (exp)\ then\ \{ Block_1 \} \ else\ \{ Block_2 \},\\...,\sigma_n,\Omega_n,Prog_n,G) \\\\-> (\sigma_1,\hat{\Omega_1}',Prog_1,...,\hat{\sigma_i}',\hat{\Omega_i}',\cdot,...,\sigma_n,\hat{\Omega_n}',Prog_n,G)}
\end{gather*}

\begin{gather*}
\drule{COND2}{ \mathcal{E}[\![ (\sigma_1,\Omega_1,Prog_1,...,\sigma_i,\Omega_i,exp,...,\sigma_n,\Omega_n,Prog_n,G) ]\!] \\\\-> (\sigma_1,\hat{\Omega_1},Prog_1,...,\hat{\sigma_i},\hat{\Omega_i},\mathbf{False},...,\sigma_n,\hat{\Omega_n},Prog_n,G)\\ (\sigma_1,\hat{\Omega_1},Prog_1,...,\hat{\sigma_i},\hat{\Omega_i},Block_2,...,\sigma_n,\hat{\Omega_n},Prog_n,G) \\\\-> (\sigma_1,\hat{\Omega_1}',Prog_1,...,\hat{\sigma_i}',\hat{\Omega_i}',\cdot,...,\sigma_n,\hat{\Omega_n}',Prog_n,G) }{(\sigma_1,\Omega_1,Prog_1,...,\sigma_i,\Omega_i,if\ (exp)\ then\ \{ Block_1 \} \ else\ \{ Block_2 \},\\...,\sigma_n,\Omega_n,Prog_n,G) \\\\-> (\sigma_1,\hat{\Omega_1}',Prog_1,...,\hat{\sigma_i}',\hat{\Omega_i}',\cdot,...,\sigma_n,\hat{\Omega_n}',Prog_n,G)}
\end{gather*}

\begin{gather*}
\drule{WHILE1}{ \mathcal{E}[\![ (\sigma_1,\Omega_1,Prog_1,...,\sigma_i,\Omega_i,exp,...,\sigma_n,\Omega_n,Prog_n,G) ]\!] \\\\-> (\sigma_1,\hat{\Omega_1},Prog_1,...,\hat{\sigma_i},\hat{\Omega_i},\mathbf{False},...,\sigma_n,\hat{\Omega_n},Prog_n,G)}{(\sigma_1,\Omega_1,Prog_1,...,\sigma_i,\Omega_i,while\ (exp)\ \{ Block \} ,...,\sigma_n,\Omega_n,Prog_n,G) \\\\-> (\sigma_1,\hat{\Omega_1},Prog_1,...,\hat{\sigma_i},\hat{\Omega_i},\cdot,...,\sigma_n,\hat{\Omega_n},Prog_n,G)}
\end{gather*}

\begin{gather*}
\drule{WHILE2}{ \mathcal{E}[\![ (\sigma_1,\Omega_1,Prog_1,...,\sigma_i,\Omega_i,exp,...,\sigma_n,\Omega_n,Prog_n,G) ]\!] \\\\-> (\sigma_1,\hat{\Omega_1},Prog_1,...,\hat{\sigma_i},\hat{\Omega_i},\mathbf{True},...,\sigma_n,\hat{\Omega_n},Prog_n,G)\\ (\sigma_1,\hat{\Omega_1},Prog_1,...,\hat{\sigma_i},\hat{\Omega_i},Block,...,\sigma_n,\hat{\Omega_n},Prog_n,G) \\\\-> (\sigma_1,\hat{\Omega_1}',Prog_1,...,\hat{\sigma_i}',\hat{\Omega_i}',\cdot,...,\sigma_n,\hat{\Omega_n}',Prog_n,G) }{(\sigma_1,\Omega_1,Prog_1,...,\sigma_i,\Omega_i,while\ (exp)\ \{ Block \} ,...,\sigma_n,\Omega_n,Prog_n,G) \\\\-> (\sigma_1,\hat{\Omega_1}',Prog_1,...,\hat{\sigma_i}',\hat{\Omega_i}',while\ (exp)\ \{ Block \} ,\\...,\sigma_n,\hat{\Omega_n}',Prog_n,G)}
\end{gather*}

\section{Semantic Rules of Evaluations}

\begin{gather*}
\drule{TE}{N_{top(M_i)}(id) \ne \mathit{None} \\
[N_{top(M_i)}(id)]^{size(T_{top(M_i)}(id))}_{top(M_i)}=v }{\mathcal{E}[\![(...,\sigma_i,\Omega_i,id,...) ]\!]->(...,\sigma_i,\Omega_i,v,...)}
\end{gather*}

\begin{gather*}
\drule{SEaa}{top(M_i).scope = (address,j) \\ N_{\Psi_{i,j}}(id) \ne \mathit{None}\\
[N_{\Psi_{i,j}}(id)]^{size(T_{\Psi_{i,j}}(id))}_{\Psi_{i,j}}=v }{\mathcal{E}[\![(...,\sigma_i,\Omega_i,id,...) ]\!]->(...,\sigma_i,\Omega_i,v,...)}
\end{gather*}

\begin{gather*}
\drule{SEas}{top(M_i).scope = (address,j) \\ N_{\Psi_{i,s}}(id) \ne \mathit{None}\\
[N_{\Psi_{i,s}}(id)]^{size(T_{\Psi_{i,s}}(id))}_{\Psi_{i,s}}=v }{\mathcal{E}[\![(...,\sigma_i,\Omega_i,id,...) ]\!]->(...,\sigma_i,\Omega_i,v,...)}
\end{gather*}

\begin{gather*}
\drule{SEag}{top(M_i).scope = (address,j) \\ N_G(id) \ne \mathit{None}\\
[N_G(id)]^{size(T_G(id))}_G=v }{\mathcal{E}[\![(...,\sigma_i,\Omega_i,id,...) ]\!]->(...,\sigma_i,\Omega_i,v,...)}
\end{gather*}

\begin{gather*}
\drule{SEss}{top(M_i).scope = engine \\ N_{\Psi_{i,s}}(id) \ne \mathit{None} \\ [N_{\Psi_{i,s}}(id)]^{size(T_{\Psi_{i,s}}(id))}_{\Psi_{i,s}}=v }{\mathcal{E}[\![(...,\sigma_i,\Omega_i,id,...) ]\!]->(...,\sigma_i,\Omega_i,v,...)}
\end{gather*}

\begin{gather*}
\drule{SEsg}{top(M_i).scope = engine \\ N_G(id) \ne \mathit{None}\\
[N_G(id)]^{size(T_G(id))}_G=v }{\mathcal{E}[\![(...,\sigma_i,\Omega_i,id,...) ]\!]->(...,\sigma_i,\Omega_i,v,...)}
\end{gather*}

\begin{gather*}
\drule{SEgg}{top(M_i).scope = global \\ N_G(id) \ne \mathit{None}\\
[N_G(id)]^{size(T_G(id))}_G=v }{\mathcal{E}[\![(\sigma_1,\Omega_1,id,...,\sigma_n,\Omega_n,id,G) ]\!] ->(\sigma_1,\Omega_1,v,...,\sigma_n,\Omega_n,v,G)}
\end{gather*}

\begin{gather*}
\drule{EFaa}{\Lambda_{scope}(idf)=address \\ top(M_i).scope = (address,j)\\\\
\Lambda_{rettype}(idf) = Type \\ \Lambda_{body}(idf) = Block \\\\
\Lambda_{paraname}(idf)= (id_1,...,id_t) \\ \Lambda_{paratype}(idf) = (T_1,...,T_t)\\\\
M_i^{(1)}=push(M_i,top(M_i)) \\ top(M_i^{(1)}).scope = (address,j)\\\\
rt = new\_ID(top(M_i^{(1)})) \\ top(M_i^{(1)}).rt = rt\\\\
(\sigma_1,\Omega_1,Prog_1,...,\sigma_i^{(1)},\Omega_i,T_1\ id_1 := exp_1;\\...;T_t\ id_t := exp_t;Type\ rt; Block,...,\sigma_n,\Omega_n,Prog_n,G) \\-> (\sigma_1,\Omega_1',Prog_1,...,\sigma_i^{(2)},\Omega_i',\cdot,...,\sigma_n,\Omega_n',Prog_n,G) \\
M_i^{(3)} = pop(M_i^{(2)}) \\[N_{top(M_i^{(2)})}(rt)]^{size(Type)}_{top(M_i^{(2)})} = v} {\mathcal{E}[\![(\sigma_1,\Omega_1,Prog_1,...,\sigma_i,\Omega_i,idf(exp_1,...,exp_t),...,\sigma_n,\Omega_n,Prog_n,G)]\!] \\\\-> (\sigma_1,\Omega_1',Prog_1,...,\sigma_i^{(3)},\Omega_i',v,...,\sigma_n,\Omega_n',Prog_n,G)}
\end{gather*}

\begin{gather*}
\drule{EFas}{\Lambda_{scope}(idf)=engine \\ top(M_i).scope = (address,j)\\\\
\Lambda_{rettype}(idf) = Type \\ \Lambda_{body}(idf) = Block \\\\
\Lambda_{paraname}(idf)= (id_1,...,id_t) \\ \Lambda_{paratype}(idf) = (T_1,...,T_t)\\\\
M_i^{(1)}=push(M_i,top(M_i)) \\ top(M_i^{(1)}).scope = engine\\\\
rt = new\_ID(top(M_i^{(1)})) \\ top(M_i^{(1)}).rt = rt\\\\
(\sigma_1,\Omega_1,Prog_1,...,\sigma_i^{(1)},\Omega_i,T_1\ id_1 := exp_1;\\...;T_t\ id_t := exp_t;Type\ rt; Block,...,\sigma_n,\Omega_n,Prog_n,G) \\\\-> (\sigma_1,\Omega_1',Prog_1,...,\sigma_i^{(2)},\Omega_i',\cdot,...,\sigma_n,\Omega_n',Prog_n,G) \\\\
M_i^{(3)} = pop(M_i^{(2)}) \\[N_{top(M_i^{(2)})}(rt)]^{size(Type)}_{top(M_i^{(2)})} = v} {\mathcal{E}[\![(\sigma_1,\Omega_1,Prog_1,...,\sigma_i,\Omega_i,idf(exp_1,...,exp_t),...,\sigma_n,\Omega_n,Prog_n,G)]\!] \\\\-> (\sigma_1,\Omega_1',Prog_1,...,\sigma_i^{(3)},\Omega_i',v,...,\sigma_n,\Omega_n',Prog_n,G)}
\end{gather*}

\begin{gather*}
\drule{EFss}{\Lambda_{scope}(idf)=engine \\ top(M_i).scope = engine\\\\
\Lambda_{rettype}(idf) = Type \\ \Lambda_{body}(idf) = Block \\\\
\Lambda_{paraname}(idf)= (id_1,...,id_t) \\ \Lambda_{paratype}(idf) = (T_1,...,T_t)\\\\
M_i^{(1)}=push(M_i,top(M_i)) \\ top(M_i^{(1)}).scope = engine\\\\
rt = new\_ID(top(M_i^{(1)})) \\ top(M_i^{(1)}).rt = rt\\\\
(\sigma_1,\Omega_1,Prog_1,...,\sigma_i^{(1)},\Omega_i,T_1\ id_1 := exp_1;\\...;T_t\ id_t := exp_t;Type\ rt; Block,...,\sigma_n,\Omega_n,Prog_n,G) \\\\-> (\sigma_1,\Omega_1',Prog_1,...,\sigma_i^{(2)},\Omega_i',\cdot,...,\sigma_n,\Omega_n',Prog_n,G) \\\\
M_i^{(3)} = pop(M_i^{(2)}) \\[N_{top(M_i^{(2)})}(rt)]^{size(Type)}_{top(M_i^{(2)})} = v} {\mathcal{E}[\![(\sigma_1,\Omega_1,Prog_1,...,\sigma_i,\Omega_i,idf(exp_1,...,exp_t),...,\sigma_n,\Omega_n,Prog_n,G)]\!] \\\\-> (\sigma_1,\Omega_1',Prog_1,...,\sigma_i^{(3)},\Omega_i',v,...,\sigma_n,\Omega_n',Prog_n,G)}
\end{gather*}

\begin{gather*}
\drule{EFgg}{\Lambda_{scope}(idf)=global \\ top(M_i).scope = global ,\ 1 \le i \le n\\\\
\Lambda_{rettype}(idf) = Type \\ \Lambda_{body}(idf) = Block \\\\
\Lambda_{paraname}(idf)= (id_1,...,id_t) \\ \Lambda_{paratype}(idf) = (T_1,...,T_t)\\\\
M_i^{(1)}=push(M_i,top(M_i)) ,\ 1 \le i \le n \\ top(M_i^{(1)}).scope = global ,\ 1 \le i \le n \\\\
rt = new\_ID(top(M_i^{(1)})) ,\ 1 \le i \le n \\ top(M_i^{(1)}).rt = rt ,\ 1 \le i \le n \\\\
(\sigma_1^{(1)},\Omega_1,T_1\ id_1 := exp_1;...;T_t\ id_t := exp_t;Type\ rt; Block,...,\\\\\sigma_n^{(1)},\Omega_n,T_1\ id_1 := exp_1;...;T_t\ id_t := exp_t;Type\ rt; Block,G) \\\\-> (\sigma_1^{(2)},\Omega_1',\cdot,...,\sigma_n^{(2)},\Omega_n',\cdot,G') \\\\
[N_{top(M_i^{(2)})}(rt)]^{size(Type)}_{top(M_i^{(2)})} = v ,\ 1 \le i \le n \\
M_i^{(3)} = pop(M_i^{(2)}),\ 1 \le i \le n \\ \sigma_i^{(3)} = \sigma_i ,\ 1 \le i \le n } {\mathcal{E}[\![(\sigma_1,\Omega_1,idf(exp_1,...,exp_t),..., \sigma_n,\Omega_n,idf(exp_1,...,exp_t),G)]\!] \\-> (\sigma_1,\Omega_1',v,...,\sigma_n,\Omega_n',v,G')}
\end{gather*}

\section{Semantic Rules of Transactions}

\begin{gather*}
\drule{IFta}{\Lambda_{scope}(idf)=address \\ top(M_i).scope = \mathit{None}\\\\
\Lambda_{rettype}(idf) = \mathit{None} \\ \Lambda_{body}(idf) = Block \\\\
\Lambda_{paraname}(idf) =(id_1,...,id_t) \;\; \Lambda_{paratype}(idf) = (T_1,...,T_t)\\\\
(i,j) = get\_sender\_address() \\ top(M_i^{(1)}).scope = (address,j) \\ M_i^{(1)}=push(M_i,\{ \}) \\ 
top(M_i^{(1)}).rt = \mathit{None} \\ M_i^{(3)} = pop(M_i^{(2)})\\\\
(\sigma_1,\Omega_1,Prog_1,...,\sigma_i^{(1)},\Omega_i,T_1\ id_1 := exp_1;\\...;T_t\ id_t := exp_t;Block,...,\sigma_n,\Omega_n,Prog_n,G) \\-> (\sigma_1,\Omega_1',Prog_1,...,\sigma_i^{(2)},\Omega_i',\cdot,...,\sigma_n,\Omega_n',Prog_n,G) 
}{(\sigma_1,\Omega_1,Prog_1,...,\sigma_i,\Omega_i,idf(exp_1,...,exp_t),...,\sigma_n,\Omega_n,Prog_n,G) \\\\ -> (\sigma_1,\Omega_1',Prog_1,...,\sigma_i^{(3)},\Omega_i',\cdot,...,\sigma_n,\Omega_n',Prog_n,G)}
\end{gather*}

\begin{gather*}
\drule{IFts}{\Lambda_{scope}(idf)=engine \\ top(M_i).scope = \mathit{None}\\\\
\Lambda_{rettype}(idf) = \mathit{None} \\ \Lambda_{body}(idf) = Block \\\\
\Lambda_{paraname}(idf) =(id_1,...,id_t) \\ \Lambda_{paratype}(idf) = (T_1,...,T_t)\\\\
(i,j) = get\_sender\_address() \\ top(M_i^{(1)}).scope = engine \\\\
M_i^{(1)}=push(M_i,\{ \}) \\ top(M_i^{(1)}).rt = \mathit{None} \\ M_i^{(3)} = pop(M_i^{(2)})\\\\
(\sigma_1,\Omega_1,Prog_1,...,\sigma_i^{(1)},\Omega_i,T_1\ id_1 := exp_1;\\...;T_t\ id_t := exp_t;Block,...,\sigma_n,\Omega_n,Prog_n,G) \\-> (\sigma_1,\Omega_1',Prog_1,...,\sigma_i^{(2)},\Omega_i',\cdot,...,\sigma_n,\Omega_n',Prog_n,G) 
}{(\sigma_1,\Omega_1,Prog_1,...,\sigma_i,\Omega_i,idf(exp_1,...,exp_t),...,\sigma_n,\Omega_n,Prog_n,G) \\\\-> (\sigma_1,\Omega_1',Prog_1,...,\sigma_i^{(3)},\Omega_i',\cdot,...,\sigma_n,\Omega_n',Prog_n,G)}
\end{gather*}

\begin{gather*}
\drule{IFtg}{\Lambda_{scope}(idf)=global \\ top(M_i).scope = \mathit{None}\\\\
\Lambda_{rettype}(idf) = \mathit{None} \\ \Lambda_{body}(idf) = Block \\\\
\Lambda_{paraname}(idf) =(id_1,...,id_t) \\ \Lambda_{paratype}(idf) = (T_1,...,T_t)\\\\
M_i^{(1)}=push(M_i,\{ \}),\ 1 \le i \le n \\\\
top(M_i^{(1)}).scope = global  ,\ 1 \le i \le n \\
top(M_i^{(1)}).rt = \mathit{None} ,\ 1 \le i \le n \\
M_i^{(3)} = pop(M_i^{(2)}) ,\ 1 \le i \le n \\
\sigma_i^{(3)} = \sigma_i ,\ 1 \le i \le n\\\\
(\sigma_1^{(1)},\Omega_1,T_1\ id_1 := exp_1;...;T_t\ id_t := exp_t;Block,...,\\\\\sigma_n^{(1)},\Omega_n,T_1\ id_1 := exp_1;...;T_t\ id_t := exp_t;Block,G) \\\\-> (\sigma_1^{(2)},\Omega_1',\cdot,...,\sigma_n^{(2)},\Omega_n',\cdot,G') }{(\sigma_1,\Omega_1,idf(exp_1,...,exp_t),...,\sigma_n,\Omega_n,idf(exp_1,...,exp_t),G) \\-> (\sigma_1,\Omega_1',\cdot,...,\sigma_n,\Omega_n',\cdot,G')}
\end{gather*}

\begin{gather*}
\drule{EFta}{\Lambda_{scope}(idf)=address \\ top(M_i).scope = \mathit{None}\\\\
\Lambda_{rettype}(idf) = Type \\ \Lambda_{body}(idf) = Block \\\\
\Lambda_{paraname}(idf)= (id_1,...,id_t) \\ \Lambda_{paratype}(idf) = (T_1,...,T_t)\\ \\
(i,j) = get\_sender\_address() \\ top(M_i^{(1)}).scope = (address,j)\\\\ 
rt = new\_ID(top(M_i^{(1)})) \\ top(M_i^{(1)}).rt = rt\\\\
(\sigma_1,\Omega_1,Prog_1,...,\sigma_i^{(1)},\Omega_i,T_1\ id_1 := exp_1;\\...;T_t\ id_t := exp_t;Type\ rt; Block,...,\sigma_n,\Omega_n,Prog_n,G) \\\\-> (\sigma_1,\Omega_1',Prog_1,...,\sigma_i^{(2)},\Omega_i',\cdot,...,\sigma_n,\Omega_n',Prog_n,G) \\\\
M_i^{(1)}=push(M_i,\{ \})  \\ M_i^{(3)} = pop(M_i^{(2)}) \\[N_{top(M_i^{(2)})}(rt)]^{size(Type)}_{top(M_i^{(2)})} = v} {\mathcal{E}[\![(\sigma_1,\Omega_1,Prog_1,...,\sigma_i,\Omega_i,idf(exp_1,...,exp_t),...,\sigma_n,\Omega_n,Prog_n,G)]\!] \\\\-> (\sigma_1,\Omega_1',Prog_1,...,\sigma_i^{(3)},\Omega_i',v,...,\sigma_n,\Omega_n',Prog_n,G)}
\end{gather*}

\begin{gather*}
\drule{EFts}{\Lambda_{scope}(idf)=engine \\ top(M_i).scope = \mathit{None}\\\\
\Lambda_{rettype}(idf) = Type \\ \Lambda_{body}(idf) = Block \\\\
\Lambda_{paraname}(idf)= (id_1,...,id_t) \\ \Lambda_{paratype}(idf) = (T_1,...,T_t)\\\\
(i,j) = get\_sender\_address() \\ top(M_i^{(1)}).scope = engine\\\\
rt = new\_ID(top(M_i^{(1)})) \\ top(M_i^{(1)}).rt = rt\\\\
(\sigma_1,\Omega_1,Prog_1,...,\sigma_i^{(1)},\Omega_i,T_1\ id_1 := exp_1;\\...;T_t\ id_t := exp_t;Type\ rt; Block,...,\sigma_n,\Omega_n,Prog_n,G) \\\\-> (\sigma_1,\Omega_1',Prog_1,...,\sigma_i^{(2)},\Omega_i',\cdot,...,\sigma_n,\Omega_n',Prog_n,G) \\\\
M_i^{(1)}=push(M_i,\{ \})  \\ M_i^{(3)} = pop(M_i^{(2)}) \\[N_{top(M_i^{(2)})}(rt)]^{size(Type)}_{top(M_i^{(2)})} = v} {\mathcal{E}[\![(\sigma_1,\Omega_1,Prog_1,...,\sigma_i,\Omega_i,idf(exp_1,...,exp_t),...,\sigma_n,\Omega_n,Prog_n,G)]\!] \\\\-> (\sigma_1,\Omega_1',Prog_1,...,\sigma_i^{(3)},\Omega_i',v,...,\sigma_n,\Omega_n',Prog_n,G)}
\end{gather*}

\begin{gather*}
\drule{EFtg}{\Lambda_{scope}(idf)=global \\ top(M_i).scope = \mathit{None} ,\ 1 \le i \le n\\\\
\Lambda_{rettype}(idf) = Type \\ \Lambda_{body}(idf) = Block \\\\
\Lambda_{paraname}(idf)= (id_1,...,id_t) \\ \Lambda_{paratype}(idf) = (T_1,...,T_t)\\\\
M_i^{(1)}=push(M_i,\{ \}) ,\ 1 \le i \le n \\ top(M_i^{(1)}).scope = global ,\ 1 \le i \le n \\\\
rt = new\_ID(top(M_i^{(1)})) ,\ 1 \le i \le n \\ top(M_i^{(1)}).rt = rt ,\ 1 \le i \le n \\\\
(\sigma_1^{(1)},\Omega_1,T_1\ id_1 := exp_1;...;T_t\ id_t := exp_t;Type\ rt; Block,...,\\\\\sigma_n^{(1)},\Omega_n,T_1\ id_1 := exp_1;...;T_t\ id_t := exp_t;Type\ rt; Block,G) \\\\-> (\sigma_1^{(2)},\Omega_1',\cdot,...,\sigma_n^{(2)},\Omega_n',\cdot,G') \\\\
[N_{top(M_i^{(2)})}(rt)]^{size(Type)}_{top(M_i^{(2)})} = v ,\ 1 \le i \le n \\
M_i^{(3)} = pop(M_i^{(2)}),\ 1 \le i \le n \\ \sigma_i^{(3)} = \sigma_i ,\ 1 \le i \le n } {\mathcal{E}[\![(\sigma_1,\Omega_1,idf(exp_1,...,exp_t),...,\sigma_n,\Omega_n,idf(exp_1,...,exp_t),G)]\!] \\\\-> (\sigma_1,\Omega_1',v,...,\sigma_n,\Omega_n',v,G')}
\end{gather*}

\end{document}